\documentstyle[11pt,newpasp,twoside,epsf]{article}

\begin{document}

\title{Measuring $\Omega_b$ from the Helium Lyman-$\alpha$ Forest}
\author{James Wadsley, Craig Hogan, Scott Anderson}
\affil{University of Washington
Department of Astronomy,
Box 351580,
Seattle, WA 98195-1580}

\begin{abstract}
A new method to extract $\Omega_b$ from high redshift intergalactic
absorption is described, based on the distribution of HeII Ly$\alpha$
optical depths in the voids in the ionization zone of quasars.  A
preliminary estimate from recent HST-STIS spectra of PKS 1935-692 at
$z=3$ gives $\Omega_b\,h^2 = 0.0125^{+0.0016}_{-0.0014}$ (1-$\sigma$
statistical errors, for a $\Lambda$CDM cosmology) consistent with
other estimates.
\end{abstract}


\section{Introduction}
In the last few years it has become possible to observe details of
absorption by singly ionized helium. The observations combine new
information about the history of quasars, intergalactic gas, and
structure formation.  

Early observations of the HeII Ly$\alpha$ absorption spectral
region included the quasars Q0302-003 (z=3.285, Jackobsen et al. 1994,
HS 1700+64 (z=2.72, Davidsen et al. 1996) and PKS 1935-692 (z=3.18
Tytler \& Jackobsen 1996).  Higher resolution (GHRS) observations of
Q0302-003 Hogan, Anderson \& Rugers 1997 and HE 2347-4342 (z=2.885,
Reimers et al. 1997) revealed structure in the absorption which could
be reliably correlated with HI absorption.  Heap et al. have followed
up with STIS on Q0302-003 (1999a) and HE 2347-4342 (1999b).  The
second observation should be particularly illuminating since HE
2347-4342 is relatively bright allowing a high resolution grating to
be used.  The Anderson et al. (1998) observations of PKS 1935-692 with
STIS yield good zero level estimates important for
estimating the optical depth $\tau$.  Preliminary reductions of longer
STIS integrations of PKS 1935-692 with the 0.1 \AA\ slit by Anderson et
al. (1999) confirm the 1998 results.  Taken together, these data now
appear to be showing the cosmic ionization of helium by quasars
around redshift 3.  Although it is possible that the medium is already ionized
to HeIII by other sources at a lower level (Miralda-Escud\'e et al. 1999).

All of the objects show absorption with mean $\tau {{
{\raisebox{-.9ex}{$>$}}\atop{\sim} }} 1$ at redshifts lower than the
quasar.  For the higher redshift QSO's Q0302-033 and PKS 1935-692
(shown in Fig. 1) there is a clear shelf of $\tau {{
{\raisebox{-.9ex}{$>$}}\atop{\sim} }} 1.3$ in a wavelength region of
order 20 {\AA} in observed wavelength blueward of the quasar emission
line redshift, dropping to a level consistent with zero flux or $\tau
{{ {\raisebox{-.9ex}{$>$}}\atop{\sim} }} 3$ beyond that.  The sharp
edge led Anderson et al. to conclude that gas initially containing
helium as mostly HeII was being double-ionized in a region around the
quasars.  The lack of a strong emission line for HeII Ly$\alpha$
suggests that ionizing flux is escaping so that the 228 {\AA} flux may
be similar to a simple power-law extension of the observed 304 {\AA}
rest frame flux.  Hogan et al. (1997) used the same reasoning to
estimate the time required for quasars to create the double ionized
helium region to be 20 Myr for a 20 {\AA} shelf (dependent on the
Hubble parameter, spectral hardness, cosmology, baryon density and the
shelf size).

The features present in HeII Ly$\alpha$ spectra are reflected in the
HI Ly$\alpha$ forest for these quasars.  Attempts to model the HeII
absorption with line systems detected in HI suggest that very low
column HI absorbers, difficult to differentiate from noise in HI
spectra, provide a substantial contribution to the HeII absorption.
Typically, in the shelf region, the ratio of HeII to HI ions is of
order 20 or more, rising to at least 100 farther away (The
cross-section for HeII Ly$\alpha$ absorption is 1/4 that of HI).  Both
PKS 1935-692 and HE 2347-4342 display conspicuous voids in the
HeII absorption near the apparent edge of the HeIII bubble with
corresponding voids in the HI spectra.

Reionization and the origin of the HeII Ly$\alpha$ forest can
addressed with detailed theoretical treatments (eg. Zheng \& Davidsen
1995, Zhang et al. 1998, Fardal et
al. 1998, Abel \& Haehnelt 1999, Gnedin 1999).  Wadsley, Hogan \&
Anderson (1999) presented numerical models of the onset of full helium
reionization around a single quasar.  The models integrated
one-dimensional radiative transfer along lines of sight taken from
cosmological hydrodynamical simulations and used flux levels
comparable to PKS 1935-692.  The results reinforced the basic
intepretation of PKS 1935-692 by Hogan et al. (1997) as the growth of
an He III bubble over time in a medium that was mostly HeII.  In
addition void recoveries in the large ionized bubbles similar to that
observed for PKS 1935-692 occurred often among a random set of
simulated lines of sight.  The gas temperature in such bubbles is
strongly affected by the ionization of He II to He III, particularly
because the first photons to reach much of the gas will be quite hard
since the softer photons are absorbed close to the quasar until the
gas is optically thin.  In particular the underdense medium reaches
temperatures of order 15000K.  We make a simple analytical argument
for this result in the next section.

Given our fairly good guess as to the physical conditions near PKS
1935-692 we are in position to attempt to convert the observed optical
depths into a density in baryons.  Comparing optical depths measured
for the small sample of voids in the shelf region of PKS 1935-692 to
distributions of void widths and densities from simulations we can
build an estimate of $\Omega_b$, the total cosmic density in baryons.
The are still uncontrollable systematic uncertainties in this estimate
but these differ from other techniques and can be addressed with
better simulations and a larger sample of quasars.  The main focus of
this paper is the technique.

\section{Estimating $\Omega_b$ Using Helium Quasar Observations}

In He II Ly$\alpha$ absorption spectra only the voids are
sufficiently low density to allow measurements of the optical depth.
The highly ionized void gas is optically thin to the ionizing
radiation and cool enough to ignore collisional effects, resulting in a fairly
simple relation between optical depth and gas density,

{\samepage{
\begin{eqnarray}
\tau_{\rm HeII}\ = 1.0 \times 10^{10} n_{\rm
HeII}/(\frac{dv}{dl}/100\  { \rm km\,s^{-1}\,Mpc^{-1}})\ \nonumber\\
 = 106\ \frac{\alpha(T)}{\Gamma}
(\rho/\bar{\rho})^2 \Omega_b h^2 \left(\frac{1+z}{4}\right)^6 \left(\frac{dv}{dl}/100\  { \rm km\,s^{-1}\,Mpc^{-1}}\right)^{-1},
\label{opacity}
\end{eqnarray}
}}

\noindent where $\tau_{\rm HeII}$ is the optical depth for absorption of He II
Ly$\alpha$, $n_{\rm HeII}$ is the number density of He II ions,
$\frac{dv}{dl}$ converts from real space to wavelength (velocity)
space and $\rho$ is the gas density.  The recombination rate
$\alpha(T)$ contains the only temperature dependence.  The
photo-ionizations per second $\Gamma$ is determined from the rest
frame HeII ionizing flux, $F_{228\AA}$ can be extrapolated from
the rest frame flux at $304 \AA$ which is estimated (assuming a
cosmological model) from the observed continuum flux at $304 \AA
(1+z_q)$, where $z_q$ is the redshift of the quasar.

The fraction of HeII depends on temperature roughly as $T^{-1/2}$
through the recombination coefficient.  The radiative transfer models
of Wadsley et al. (1999) gave temperatures around 15000K for the
underdense gas near quasars.  The optical depth to a He II ionizing
photon travelling $\delta z = 0.1$ (The size of the PKS 1935-692 shelf
region) at redshift $z=3$ (SCDM, $\Omega_b=0.05$, $H_0 = 50
{\rm km\,s^{-1}\,Mpc^{-1}}$) is approximately $\tau = 62
(E_{\gamma}/54.4{\rm\,eV})^{-3}$.  The optical depth falls to 1 for
photons with energies, $E_{\gamma}\sim 215{\rm\,eV}$.  Thus the HeII
in the outer half of the shelf region will be ionized preferentially
by photons in the energy range 100-200 eV.  Injecting 100-54.4 eV per
helium atom is equivalent to a temperature increase of 13000 K when
the energy is distributed among all the particles.  The cooling time
for the underdense gas is very long, dominated by adiabatic expansion.
Gas elsewhere will gain $\sim$ 4000 K due to HeIII reionization.

\begin{figure}[t]
\plotone{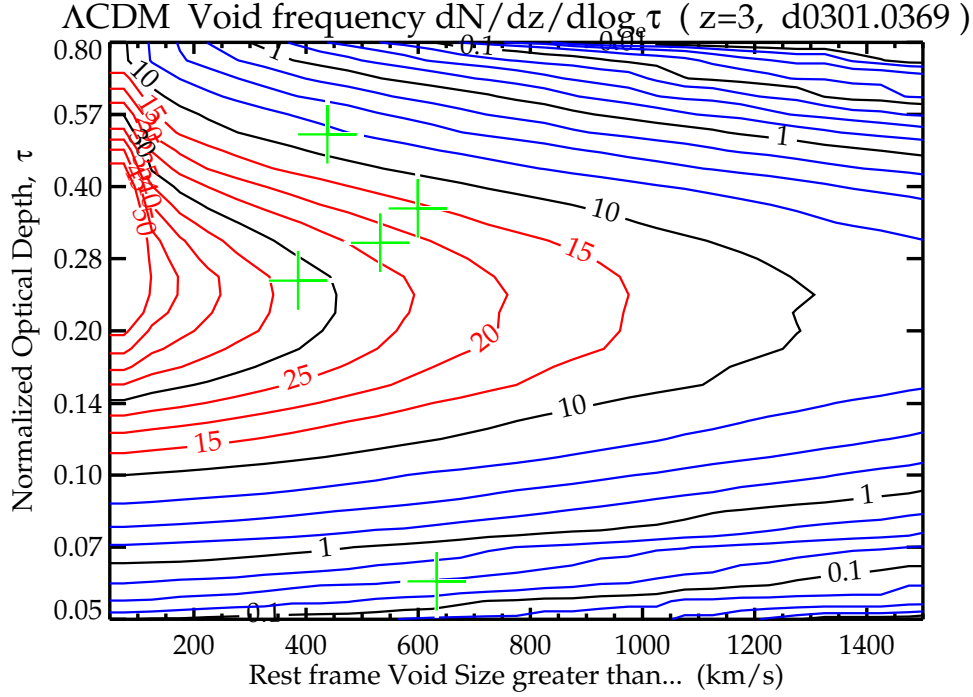}
\vspace{-0.1in}
\caption{Probability contours for void optical depths and velocity widths
along random lines of sight in a $\Lambda$CDM simulation with the 5
void optical depths and widths from PKS 1935-692 overplotted as green
crosses.  The opticals depths (in helium II Lyman-$\alpha$ absorption) are normalized so that mean density gas expanding with the Hubble flow at $z=3$ has an optical depth of one.
The probabilities are cumulative in velocity and are
approximately indicative of the number of voids to be expected per
unit redshift interval.  The simulations (described in the text) predict $\sim 3$ voids with sizes $\ga$ 300 ${\rm km\,s^{-1}}$ in the interval $z=3.09$ to $z=3.15$ where 5 were detected.}

\label{taunorm}
\end{figure}

As voids evolve they approach an attractor solution resembling empty
universes and their relative growth slows when they reach around $0.1$
times the mean density, making that value fairly representative.  We
used $N$-body simulations of 3 cosmologies (Standard CDM, Open CDM and
$\Lambda$CDM) to get probability distributions for the quantity we
label normalized opacity
$O=(\rho/\bar{\rho})^2/(\frac{dv}{dl}/\frac{dv}{dl}_{HUBBLE})$ as a
function of void width.  $O$ may be directly related to the void
optical depths via (\ref{opacity}).  The set of voids allows a
statistical maximum likelihood comparison of the simulated
distribution of $O$ to the distribution of void widths and optical
depths observed.  Figure~\ref{taunorm} shows the probability
distribution of $O$ in voids (equivalent to void optical depth) and
void widths for the $\Lambda$CDM cosmology.  The optical depth from
the simulation was normalized so that mean density gas expanding with
the Hubble flow gives an optical depth of 1.0.

The effect of gas pressure on the dynamics of gas in the voids should
be negligible and was ignored in our simulations.  Thus the gas
follows the dark matter exactly.  The simulations were performed using
Hydra (Couchman 1991) with 64 Mpc periodic simulation volumes ($\sim
6000$ ${\rm km\,s^{-1}}$ at $z=3$) containing $128^3$ particles.  A
preliminary convergence study indicated that the statistics of $O$ are
relatively insensitive to resolution however our resolution is still
lower than that of the best current Lyman-$\alpha$ forest studies which also
include gas.  We were constrained by the need to include larger scales
so that the large void portion of our sample was reasonable.  Higher
resolution gives slightly larger estimates for $\Omega_b$.

The ratio of $\tau_{\rm HeII}$ to $O$ that has the maximum likelihood gives an estimate for the total baryon density as follows,
\begin{eqnarray}
{\Omega_b\,h^2} = 0.0178 \left( \frac{\tau_{\rm HeII}}{O}/{\ 3.5}
\right)^{\frac{1}{2}} {\left(
\frac{{F_{\nu}}_{\, {228\AA},\, {\rm REST}}}{6\times 10^{-23}
erg\,s^{-1}Hz^{-1}cm^{-2}} \right)}^{\frac{1}{2}}
\nonumber \\ \left( \frac{H(z)}{400{\rm
km\,s^{-1}\,Mpc^{-1}}} \right) ^{\frac{1}{2}}  {\left(
\frac{3+\alpha}{4.8} \right) }^{-\frac{1}{2}} {\left(
\frac{Y}{0.25} \right) }^{\frac{1}{2}} {\left( \frac{1+z}{4.0} \right)
}^{-3} {\left( \frac{T}{15000\,K} \right) } ^{\frac{1}{4}}.
\label{ombh2eqn}
\end{eqnarray}
In this equation all the quantities are given as ratios with typical
values estimated from PKS 1935-692, with ${\rho/\bar{\rho}}$ being the
local density divided by the cosmological mean density,
{$\frac{dv}{dl}/{\frac{dv}{dl}}_{\rm HUBBLE}$} the local expansion
rate along the line-of-sight compared to the mean, $H(z)=H_0\,
(\Omega_M (1+z)^3 + \Omega_{\rm CURV} (1+z)^2 +
\Omega_{\Lambda})^{1/2}$ local Hubble parameter value, {$\alpha$} the
spectral slope of the quasar ($F_{\nu}\propto \nu^{-\alpha}$) and $Y$
the helium mass fraction.  The value of $H(z)$ is $400\
{\rm km\,s^{-1}\,Mpc^{-1}}$ at $z=3$ for an $H_0 = 50\ {\rm km\,s^{-1}\,Mpc^{-1}}$
SCDM cosmology.

\begin{figure}[t]
\plotone{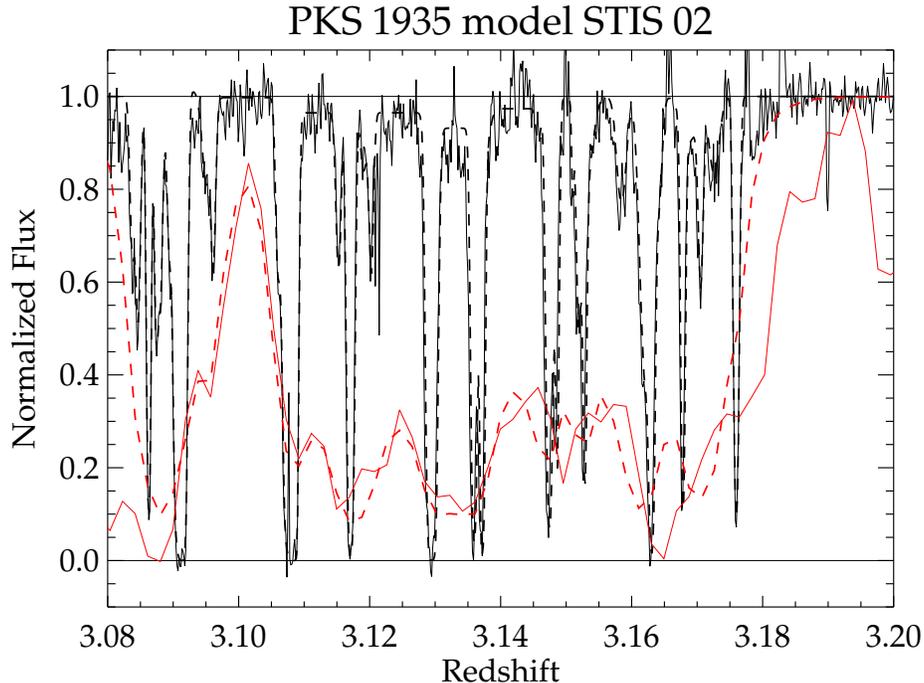}
\vspace{-0.1in}
\caption{Data and models of PKS-1935-692 HST/STIS Data. The top (black) curves are ground based HI data (solid) and the model (dashed).  The lower (red) curves are HeII data from STIS (solid) and the model (dashed).  Non-zero optical depth in 5 voids was required to fit the data between $z=3.09$ and $z=3.15$: a region dominated by the Quasar UV Flux.  This region should be free of associated absorbers.  }
\label{taufit}
\end{figure}

\section{Results for PKS 1935-692}

The exact systemic redshift of PKS 1935-692 is uncertain due to an absence
of narrow IR emission lines.  We use $z=3.19$.

We used the 5 large voids observed over the range $z=3.09-3.15$ for
PKS~1935-692. This range should be far enough from the quasar to avoid
``associated'' absorbers caught in the quasar outflow.  The fits are
shown in figure~\ref{taufit}. He II and H I absorption are both
plotted with the models for each overplotted as dashed lines.  The He II
spectrum was modeled with the known HI Ly$\alpha$ absorption lines
and 5 parameter values for the optical depth in each void.  The
ratio of the HI to HeII optical depths was a sixth free parameter.
The models were convolved with the same spectral point spread function
as the observations and include same level of photon shot noise.
The best model was determined using maximum likelihood and the fitting
errors determined with Monte Carlo realizations.  The emptiest void
has a reasonable probability according to the simulations, however
since it has been suggested that it could be due to a local ionizing
source we calculated the effect of removing it, which was a $\sim
40\%$ increase in the estimate for $\Omega_b$.

The ionizing flux was inferred from the observed PKS 1935-692
continuum level at 304 $\AA$ rest frame and extrapolated to $z>3$ and
228 $\AA$.  Assuming the quasar redshift is $z=3.19$ this gives a flux
of $6\times 10^{-23} erg\,s^{-1}Hz^{-1}cm^{-2}$ (for standard CDM) at
$z=3.125$ which is in the middle of the voids.  The sample of helium
quasars is quite small and though PKS 1935-692 does not have any known
variability all quasars are thought to be intrinsically variable due
to the nature of the fueling and emission processes.  Selection
effects favour the idea that PKS 1935-692 is currently brighter than
its average value over the last few thousand years (the ionization
response timescale).  If PKS 1935-692 has recently brightened then
(~\ref{ombh2eqn}) implies that our $\Omega_b$ estimate is biased
upward.  If there is additional foreground continuum absorption then
our estimates are biased low.  

The results are tabulated below with 1-$\sigma$ fitting errors indicated
(Monte Carlo).
Aside from the uncertainties mentioned above, there is still
uncertainty in the temperature through the $T^{1/4}$ dependence.  If
the gas was pre-ionized to HeIII the underdense gas could be colder by
a factor of order 2 which would lower the $\Omega_b h^2$ estimates
by 20 \%.  A larger sample of quasars would make the greatest
improvement in the robustness of this measurement.

\begin{table}
\begin{tabular}{|lllll|l|}
\hline
 & $\Omega_M$ & $\Omega_{\Lambda}$ & $H_0$ & $\sigma_8$ & {$\Omega_b\,h^2$} \\
 & & & {\scriptsize ${\rm km\,s^{-1}\,Mpc^{-1}}$} & & \\
\hline
{$\Lambda$ CDM} & 0.335 & 0.665 & 70 & 0.91 & 0.0125 $^{+0.0016}_{-0.0014}$ \\

{Open CDM} & 0.37 & & 70 & 0.91 & 0.0186 $^{+0.0026}_{-0.0023}$ \\

{Standard CDM} & 1.0 & & 50 & 0.67 & 0.0150 $^{+0.0021}_{-0.0019}$ \\
\hline
\end{tabular}
\caption{ Preliminary results for $\Omega_b\,h^2$ for 3 Cosmologies.  The errors shown are fitting errors: systematic uncertainty is significantly greater.}
\end{table}

\end{document}